\documentclass[pre,twocolumn,aps]{revtex4}
\newcommand{\bec}[1]{\mbox{\boldmath $ #1$}}
\usepackage{graphicx}
\begin{document}
\title{Hysteresis phenomenon in turbulent convection}

\author{A.~Eidelman}
\author{T.~Elperin}
\email{elperin@bgu.ac.il}
\author{N.~Kleeorin}
\author{A.~Markovich}
\author{I.~Rogachevskii}
\affiliation{The Pearlstone Center for Aeronautical Engineering
Studies, Department of Mechanical Engineering, Ben-Gurion University
of the Negev, Beer-Sheva 84105, P. O. Box 653, Israel}
\date{\today}
\begin{abstract}
Coherent large-scale circulations of turbulent thermal convection in
air have been studied experimentally in a rectangular box heated
from below and cooled from above using Particle Image Velocimetry.
The hysteresis phenomenon in turbulent convection was found by
varying the temperature difference between the bottom and the top
walls of the chamber (the Rayleigh number was changed within the
range of $10^7 - 10^8$). The hysteresis loop comprises the one-cell
and two-cells flow patterns while the aspect ratio is kept constant
($A=2 - 2.23$). We found that the change of the sign of the degree
of the anisotropy of turbulence was accompanied by the change of the
flow pattern. The developed theory of coherent structures in
turbulent convection (Elperin et al. 2002; 2005) is in agreement
with the experimental observations. The observed coherent structures
are superimposed on a small-scale turbulent convection. The
redistribution of the turbulent heat flux plays a crucial role in
the formation of coherent large-scale circulations in turbulent
convection.
\end{abstract}

\maketitle

\section{Introduction}

Coherent structures in turbulent convection are observed in the
atmospheric convective boundary layers (see, e.g., Etling and Brown
1993; Atkinson and Wu Zhang 1996; Br\"{u}mmer 1999) and in
laboratory experiments (see, e.g., Krishnamurti and Howard 1981;
Siggia 1994; Kadanoff 2001). In the atmospheric shear-free
convection, the coherent structures represent large-scale
three-dimensional long-lived B{\'e}nard-type cells (cloud cells)
composed of narrow uprising plumes and wide downdraughts. They
usually embrace the entire convective boundary layer (of the order
of 1-3 km in height) and include pronounced convergence flow
patterns close to the surface. In the sheared convective flows, the
coherent structures represent convective boundary layer scale rolls
(cloud streets) stretched along the mean wind (see, e.g.,  Etling
and Brown 1993; Atkinson and Wu Zhang 1996; Br\"{u}mmer 1999).

Buoyancy-driven structures, such as plumes, jets, and large-scale
circulation patterns were observed in numerous laboratory
experiments. The large-scale circulation caused by convection in the
Rayleigh-B{\'e}nard apparatus is often called the "mean wind" (see,
e.g., Krishnamurti and Howard 1981; Zocchi et al. 1990; Ciliberto et
al. 1996; Niemela et al. 2001; Sreenivasan et al. 2002; Niemela and
Sreenivasan 2003; Burr et al. 2003; Shang et al. 2004; Xi et al.
2004; Tsuji et al. 2005; Brown et al. 2005, and references therein).
There are several open questions concerning these flows, e.g., how
do they arise, and what are their characteristics and dynamics.

The life-times of coherent structures are very long compared to the
largest turbulent time-scales. Their spectral properties differ from
those of small-scale turbulence. They are characterized by narrow
spectra and do not exhibit the direct energy-cascade behavior (from
larger to smaller scales). As a result the turbulence and the
coherent structures interact in practically the same way as the
turbulence and the mean flow. These structures show more similarity
in their behavior with regular flows than with turbulence. They can
be identified as the motions, whose spatial and temporal scales are
much larger than the characteristic turbulent scales.

In spite of a number of theoretical and numerical studies (see,
e.g., Busse and Whitehead 1971; 1974; Busse 1983; Lenschow and
Stephens 1980; Hunt 1984; Hunt et al. 1988; Schmidt and Schumann
1989; Zilitinkevich 1991; Williams and Hacker 1992; 1993;
Zilitinkevich et al. 1998; Young et al. 2002; Parodi et al. 2004,
and references therein), the nature of large-scale coherent
structures in turbulent convection is a subject of discussions.
Hartlep et al. (2003) noted that there are two points of view on the
origin of large-scale circulation in turbulent convection.
"According to one point of view, the rolls which develop at low
Rayleigh numbers, ${\rm Ra}$, near the onset of convection
continually increase their size as ${\rm Ra}$ is increased and
continue to exist in an average sense at even the highest Rayleigh
numbers reached in the experiments (see, e.g., Fitzjarrald 1976).
Another hypothesis holds that the large-scale circulation is a
genuine high Rayleigh number effect (see, e.g., Krishnamurti and
Howard 1981)."

The new mechanism of formation of the large-scale coherent
structures in turbulent convection was proposed recently by Elperin
et al. (2002; 2005). It was suggested that the redistribution of the
turbulent heat flux plays a crucial role in the formation of the
large-scale circulations in turbulent convection. In particular, two
competitive effects, namely redistribution of the vertical turbulent
heat flux due to convergence or divergence of the horizontal mean
flows, and production of the horizontal component of the  turbulent
heat flux due to the interaction of the mean vorticity with the
vertical component of the  turbulent heat flux, cause the
large-scale instability and formation of the large-scale coherent
structures in turbulent convection (Elperin et al. 2002; 2005).

The main goal of this paper is to describe the experimental study of
large-scale circulations of turbulent thermal convection in air flow
(the aspect ratios $A=2 - 2.23$). In order to study large-scale
circulations we used Particle Image Velocimetry to determine the
turbulent and mean velocity fields, and a specially designed
temperature probe with twelve sensitive thermocouples was employed
to measure the temperature field. We found the hysteresis phenomenon
in turbulent convection by varying the temperature difference
between the bottom and the top walls of the chamber (the Rayleigh
number was changed within the range of ${\rm Ra} = 10^7 - 10^8$).
The hysteresis loop comprises the one-cell and two-cell flow
patterns. We found that the change of the sign of the degree of the
anisotropy of turbulence was accompanied by the change of the
pattern of the mean wind. The developed theory of coherent
structures in turbulent convection (Elperin et al. 2002; 2005) is in
a good agreement with the experimental observations.

The hysteresis phenomenon in laminar convection was found by Busse
(1967) who defined it as follows: "The fact that the convection at a
certain Rayleigh number depends on the way in which the Rayleigh
number has been reached is called the hysteresis effect in laminar
convection". In the laminar convection the hexagon flow structures
transform into rolls structures by increasing the Rayleigh number.
On the other hand, decreasing the Rayleigh number causes the
transition from the rolls structure to the hexagons. Similar
phenomena were observed in numerical simulations and laboratory
experiments (see, e.g., Busse 1978; Braunsfurth et al. 1996; Gelfgat
et al. 1999, and references therein).

In the experiments by Willis et al. (1972) performed in air, water
and a silicon oil for the Rayleigh numbers ${\rm Ra} = (0.2 - 3)
\times 10^4$ and large aspect ratios (for air $A=31.5$), it was
demonstrated that the average dimensionless roll diameter,
$\Lambda$, increases as ${\rm Ra}$ is increased. This observation
was less pronounced for large Rayleigh number flows (for water and
silicon oil). The hysteresis phenomenon was found in dependence
$\Lambda({\rm Ra})$ for large Prandtl number flows (for water and
silicon oil), but this phenomenon was not observed in the air flow
(Willis et al. 1972). For considerably larger values of the Rayleigh
numbers the convection patterns become very complicated (Willis et
al. 1972) and turbulent convection arises. In the experiments by
Fitzjarrald (1976) in air flow it was shown that in convection with
large aspect ratios the predominant horizontal scale of the regular
structures increased from $4 h$ at ${\rm Ra} = 10^5$ to $6 h$ at
${\rm Ra} = 10^6$ (where $h$ is the vertical distance between
plates). In the range ${\rm Ra} = 10^6-10^7$ this scale did not
change (Fitzjarrald 1976). Notably, the hysteresis phenomenon was
not observed by Fitzjarrald (1976).

In the direct numerical simulations of convection by Hartlep et al.
(2003) for the Rayleigh numbers ${\rm Ra} = 10^3 - 10^6$, the
Prandtl number $\, {\rm Pr}=0.7$ and large aspect ratio ($A=10$) it
was found that "the typical size of the large-scale structures does
not always vary monotonically as a function of ${\rm Ra}$, but
broadly increases with increasing ${\rm Ra}$. It cannot be decided
from these simulations whether the large-scale structures will
eventually disappear at yet higher Rayleigh numbers." Notably, the
hysteresis phenomenon was not observed by Hartlep et al. (2003).

The paper is organized as follows. Section 2 discusses the physics
of formation of the large-scale coherent structures, and Section 3
describes the experimental set-up for a laboratory study of this
effect. The experimental results and their detailed analysis are
presented in Section 4. Finally, conclusions are drawn in Section
5.

\section{Turbulent heat flux and coherent structures}

Traditional theoretical models of the boundary-layer turbulence,
such as the Kolmogorov-type local closures, imply the following
assumptions. Fluid flows are decomposed into two components of
principally different nature, organized mean flow and turbulent
flow. Turbulent fluxes are proportional to the local mean
gradients, whereas the proportionality coefficients (eddy
viscosity, turbulent diffusivity) are uniquely determined by local
turbulent parameters. For example, a widely used traditional
approximation for the turbulent heat flux reads ${\bf F} = \langle
s {\bf u} \rangle = - \kappa_T \bec{\nabla} S $ (see, e.g., Monin
\& Yaglom 1975), where $\kappa_T$  is the turbulent thermal
conductivity, $S$ is the mean entropy, ${\bf u}$ and $s$ are
fluctuations of the velocity and entropy, respectively.

The traditional form for the turbulent heat flux ${\bf F}$ does not
include the contribution from anisotropic velocity fluctuations.
Actually the mean velocity gradients can directly affect the
turbulent heat flux. The reason is that additional essentially
anisotropic velocity fluctuations are generated by tangling the
mean-velocity gradients with the Kolmogorov-type turbulence due to
the influence of the inertial forces during the life time of large
turbulent eddies. Therefore, the Kolmogorov turbulence supplies
energy to the anisotropic (tangling) turbulence. In its turn the
anisotropic turbulence causes formation of coherent structures due
to the excitation of a large-scale instability (Elperin et al. 2002;
2005). Anisotropic velocity fluctuations were studied for the first
time by Lumley (1967). He had shown that the velocity field in the
presence of mean shear is strongly anisotropic and is characterized
by a steeper spectrum than the Kolmogorov turbulence (see also
Wyngaard and Cote 1972; Saddoughi et al. 1994; Ishihara et al. 2002;
Yoshida et al. 2003, among others).

In order to parameterize the anisotropic turbulence, a spectral
closure model was developed by Elperin et al. (2002; 2005). The
derivation of the anisotropic turbulence model includes the
following steps: applying the spectral closure, solving the
equations for the second moments in the ${\bf k}$ space, and
returning to the physical space to obtain formulas for the Reynolds
stresses and the turbulent heat flux. The derivation are based on
the Navier-Stokes equation and the entropy evolution equation
formulated in the Boussinesq approximation. This derivation yields
the following expression for the turbulent flux of entropy ${\bf F}
\equiv \langle s {\bf u} \rangle $:
\begin{eqnarray}
{\bf F} = {\bf F}^{\ast} -  \tau \, \biggl[\alpha \, {\bf
F}_{z}^{\ast} \, {\rm div} \, {\bf U}_{h} - {\tau \over 5} \,
\biggl(\alpha + {3 \over 2} \biggr) \, ({\bf W} {\bf \times} {\bf
F}_z^{\ast}) \biggr] \; \label{A1}
\end{eqnarray}
(for details, see Elperin et al. 2002; 2005), where $\tau$ is the
correlation time of the Kolmogorov turbulence corresponding to the
maximum scale of turbulent motions, ${\bf W} = \bec{\nabla} {\bf
\times} {\bf U}$ is the mean vorticity, ${\bf U} = {\bf U}_{h} +
{\bf U}_{z}$ is the mean velocity with the horizontal ${\bf U}_{h}$
and vertical ${\bf U}_{z}$ components, $\alpha$ is the degree of
thermal anisotropy, $F_i^{\ast} = - \kappa_{ij} {\nabla}_j S $ is
the background turbulent heat flux,
\begin{eqnarray}
\kappa_{ij} = \kappa_{T} \biggl[\delta_{ij} + {3\over 2} (2+\tilde
\gamma) e_i e_j \biggr] \label{A3}
\end{eqnarray}
is a generalized anisotropic turbulent heat conductivity tensor,
$\tilde \gamma$ is the ratio of specific heats and ${\bf e}$ is
the vertical unit vector. The equation for the tensor
$\kappa_{ij}$ was derived in Appendix A in Elperin et al. (2002)
using the budget equations for the turbulent kinetic energy,
fluctuations of the entropy and the turbulent heat flux. The
anisotropic part of the tensor $\kappa_{ij}$ (the second term in
the square brackets) is caused by a modification of the turbulent
heat flux by the buoyancy effects. The parameter $\alpha$ is given
by
\begin{eqnarray}
\alpha &=& {1 + 4 \xi \over 1 + \xi / 3} \;, \quad  \quad \xi  =
\biggl({l_{h} \over l_{z}} \biggr)^{2/3} - 1 \;,
\label{A4}
\end{eqnarray}
where $l_{h}$ and $l_{z}$ are the horizontal and vertical scales
in which the background turbulent heat flux $F_{z}^{\ast}({\bf r})
= \langle s({\bf x}) \, u_z({\bf x}+{\bf r}) \rangle$ tends to
zero. The parameter $\xi$ describes the degree of thermal
anisotropy. In particular, when $l_{h} = l_{z}$ the parameter $\xi
= 0$ and $\alpha = 1$. For $l_{h} \ll l_{z}$ the parameter $\xi =
- 1$ and $\alpha = - 9/2$. The maximum value $ \xi_{\rm max} $ of
the parameter $\xi$ is given by $\xi_{\rm max} = 2/3$ for $\alpha
= 3$. The upper limit for the parameter $\xi$ arises because the
function $F_{z}^{\ast}({\bf r})$ has a global maximum at ${\bf
r}=0$. Thus, for $\alpha < 1$ the thermal structures have the form
of columns or thermal jets $(l_{h} < l_{z})$, and for $\alpha > 1$
there exist the `'pancake'' thermal structures $(l_{h} > l_{z})$
in the background turbulent convection (i.e., a turbulent
convection with a zero gradients of the mean velocity).

The terms in the square brackets in the right hand side of
Eq.~(\ref{A1}) are caused by the anisotropic turbulence and depend
on the "mean" (including coherent) velocity gradients. These terms
lead to the excitation of large-scale instability and formation of
coherent structures. In Eq.~(\ref{A1}) the terms with zero
divergence are omitted, because only ${\rm div} \, {\bf F}$
contributes to the mean-field dynamics. Neglecting the anisotropic
turbulence term this equation reduces to the traditional equation.
The physical meaning of Eq.~(\ref{A1}) is the following. The second
term in Eq.~(\ref{A1}) describes the redistribution of the vertical
background turbulent heat flux by convergent (or divergent)
horizontal mean flows. This redistribution of the vertical turbulent
heat flux occurs during the life-time of turbulent eddies. The last
term in Eq.~(\ref{A1}) determines the formation of the horizontal
turbulent heat flux due to "rotation" of the vertical background
turbulent heat flux by the perturbations of the horizontal mean
vorticity. The emergence of the horizontal turbulent heat flux is
caused by the action of the local inertial forces in inhomogeneous
mean flows.

\begin{figure}
\vspace*{2mm} \centering
\includegraphics[width=8cm]{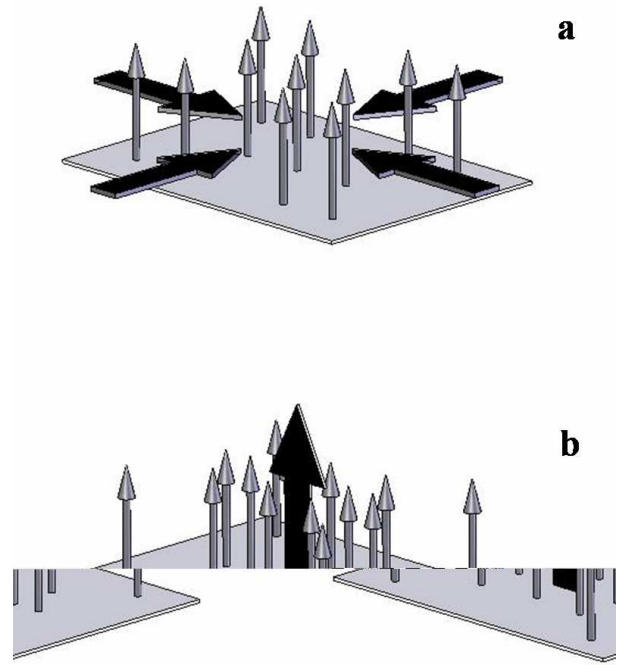}
\caption{\label{Fig1} Effect of a nonzero ${\rm div} \, {\bf
U}_{h}$ which causes a redistribution of the vertical turbulent
flux of the entropy and results in a formation of a large-scale
circulation of the velocity field:  (a) fluid flow with $ {\rm
div} \, {\bf U}_{h} < 0 $ produces regions with enhanced vertical
fluxes of entropy and vertical fluid flow in these regions (b).}
\end{figure}

In the shear-free regime, the large-scale instability is related to
the first term in square brackets in Eq.~(\ref{A1}) for the
turbulent flux of entropy. When $\partial U_{z} /\partial z > 0$,
perturbations of the vertical velocity $U_{z}$ cause negative
divergence of the horizontal velocity, ${\rm div} \, {\bf U}_{h} < 0
$ (provided that ${\rm div} \, {\bf U} = 0). $ This strengthens the
local vertical turbulent flux of entropy and causes increase of the
local "mean" entropy and buoyancy (see Fig. 1). The latter enhances
the local "mean" vertical velocity $U_{z}$. Thus a large-scale
instability is excited. Similar reasoning is valid when $\partial
U_{z} /\partial z < 0$, whereas ${\rm div} \, {\bf U}_{h} > 0 $.
Then a negative perturbation of the vertical flux of entropy leads
to a decrease of the "mean" entropy and buoyancy, that enhances the
downward flow and once again excites the instability. Therefore,
nonzero ${\rm div} \, {\bf U}_{h}$ causes redistribution of the
vertical turbulent flux of entropy and formation of regions with
large values of this flux. These regions (where ${\rm div} \, {\bf
U}_{h} < 0 $) alternate with the low-flux regions (where ${\rm div}
\, {\bf U}_{h} > 0 $). By this means large-scale coherent structures
are formed.

The role of the second term in square brackets in Eq.~(\ref{A1})
is to decrease the growth rate of the large-scale instability for
$\alpha > - 3/2$. Indeed, the interaction of the "mean" vorticity
with the vertical flux of entropy produces the horizontal
turbulent heat flux. The latter decreases (increases) the "mean"
entropy in the regions with upward (downward) local flows, thus
diminishing the buoyancy forces and reducing the "mean" vertical
velocity $U_{z}$ and the "mean" vorticity ${\bf W}$. This
mechanism dampens the large-scale instability.

The above two competitive effects determine the growth rate of the
large-scale instability. Perturbation analysis of the mean-field
equations in the shear-free convection regime results in the
following expression for the growth rate $\gamma$ of long-wave
perturbations:
\begin{eqnarray}
\gamma \propto g \, F_z^\ast \, \tau^2 K^{2} \, \sqrt{\beta} \, |
\sin \theta | \, \biggl[\alpha - {3 \over 8} - {5 \alpha \over 4}
\sin^{2} \theta \biggr]^{1/2} \; \label{A5}
\end{eqnarray}
(Elperin et al. 2002; 2005), where the parameter $\beta=(l \,
K)^{-2} \gg 1$, $\, l$  is the maximum scale of turbulent motions,
$\theta$ is the angle between the vertical unit vector ${\bf e}$ and
the wave vector ${\bf K}$ of small perturbations, ${\bf g}$ is
acceleration of gravity. Equation~(\ref{A5}) was derived for the
case $N^2 \ll g F_z^\ast \tau K^{2}$, where $N$ is the
Brunt-V\"{a}is\"{a}l\"{a} frequency. The large-scale instability
occurs when $ \alpha (5 \cos^{2} \theta - 1)
> 3/2 .$ This yields two ranges for the instability:
\begin{eqnarray}
{3 \over 2 (5 \cos^{2} \theta - 1)} < \alpha < 3 \;,
\label{A8}\\
- {9 \over 2} < \alpha < - {3 \over 2 (1 - 5 \cos^{2} \theta)} \;,
\label{A9}
\end{eqnarray}
where we took into account that the parameter $ \alpha $ varies in
the interval $ - 9 / 2 < \alpha < 3 $. The first range for the
large-scale instability in Eq.~(\ref{A8}) is for the angles $ 3/10
\leq \cos^{2} \theta \leq 1 $ (the ratio $ 0 < L_{z} / L_{h} < 1.53)
,$ and the second range for the large-scale instability (see
Eq.~(\ref{A9})) corresponds to the angles $ 0 \leq \cos^{2} \theta <
2 / 15 $ (the ratio $ 2.55 < L_{z} / L_{h} < \infty ) ,$ where $
L_{z} / L_{h} \equiv K_{h} / K_{z} = \tan \theta $ and $K_{h}$ is
the horizontal component of the wave vector. The maximum growth rate
of the large-scale instability is achieved at the scale of
perturbations $L \approx 10 \, l$, where $ L \equiv 1 /
\sqrt{L_{z}^{-2} + L_{h}^{-2}} $. The characteristic time of
excitation of the large-scale instability is of the order of $20-30
\, \tau$. Therefore, the typical length and time scales of these
structures are much larger than the turbulence scales. This
justifies the separation of scales assumption required for treating
coherent structures as albeit complex but "mean" flow. The growth
rate of the large-scale instability depends also on the degree of
anisotropy of turbulent velocity field. The detailed analysis of
this facet is performed in Section 4 in order to explain the
obtained experimental results.

\section{Experimental set-up}

Now we describe the experimental study of large-scale circulations
formed in turbulent convection. The experiments were conducted in a
rectangular chamber with dimensions $26 \times 58 \times 26$ cm in
air flow (see Fig.~2). The side walls of the chamber are made of
transparent Perspex with the thickness of $10$ mm. A number of
experiments were also conducted with different additional thermal
insulation of the side walls of the chamber in order to study
whether a heat flux through the side walls affects the turbulent
convective pattern. First, the side walls of the chamber were
insulated with Styrofoam plates with low thermal conductivity
($\kappa \sim 0.033$ W/(m K)) and with the thickness of 30 mm. Two
of the side plates were removed for a short time when the images of
the flow were recorded. In the next series of experiments we
installed additional Perspex plates with a thickness of 6 mm and
with an air gap of 2 mm between these plates and the outside walls
of the chamber. Finally, we performed experiments where these two
types of thermal insulation were used simultaneously.

One vertically oriented stationary grid (with $50$ mm mesh and $10$
mm bars where bars are arranged in a square array) was attached to
the left horizontal rod. This grid was positioned at a distance of
one mesh size from the left wall of the chamber parallel to it.
Hereafter, we use the following system of coordinates: $Z$ is the
vertical axis, the $Y$-axis is perpendicular to the grid and the
$XZ$-plane is parallel to the grid plane.

\begin{figure}
\vspace*{2mm}
\centering
\includegraphics[width=8cm]{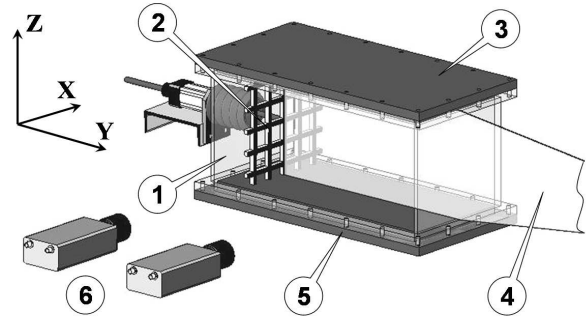}
\caption{\label{Fig2} Experimental set-up: (1) - walls of the
chamber; (2) - vertically oriented stationary grid; (3) - cooled
top with a heat exchanger; (4) - laser light sheet (5) - heated
bottom with a heat exchanger; (6) - two progressive-scan 12 bit
digital CCD cameras.}
\end{figure}

A vertical mean temperature gradient in the turbulent flow was
formed by attaching two aluminium heat exchangers to the bottom and
top walls of the test section (a heated bottom and a cooled top wall
of the chamber). The thickness of the aluminium plates is 2.5 cm.
The top plate is a bottom wall of the tank with cooling water.
Temperature of water circulating through the tank and the chiller is
kept constant within $0.1$ K. Cold water is pumped into the tank
through two inlets and flows out through two outlets located at the
side wall of the chamber. The bottom plate is attached to the
electrical heater that provides constant and uniform heating. The
voltage of a stable power supply applied to the heater varies in the
range from 35 V to 200 V. The power of the heater varies in the
range from 10 W to 300 W. The temperatures of the conducting plates
were measured with two thermocouples attached at the surface of each
plate. The temperature difference  between the top and bottom
plates, $\Delta T$,  varies in the range from 5 K to 80 K depending
on the power of the heater (i.e., the Rayleigh number was changed
within the range of ${\rm Ra} = 10^7 - 1.6 \times 10^8$). The
temperature in the probed region was measured with a specially
designed temperature probe with twelve sensitive thermocouples. The
temperature measurements showed that the thermal structure inside
the large-scale circulation is inhomogeneous and anisotropic. The
hot thermal plumes accumulate at one side of the large-scale
circulation, and cold plumes concentrate at the opposite side of the
large-scale circulation.

The velocity field was measured using a Particle Image Velocimetry
(PIV), see Raffel  et al. (1998). A digital PIV system with LaVision
Flow Master III was used. A double-pulsed light sheet was provided
by a Nd-YAG laser (Continuum Surelite $ 2 \times 170$ mJ). The light
sheet optics includes spherical and cylindrical Galilei telescopes
with tuneable divergence and adjustable focus length. We used two
progressive-scan 12 bit digital CCD cameras (with pixel size $6.7 \,
\mu$m $\times 6.7 \, \mu$m and $1280 \times 1024$ pixels) with a
dual-frame-technique for cross-correlation processing of captured
images. A programmable Timing Unit (PC interface card) generated
sequences of pulses to control the laser, camera and data
acquisition rate. The software package LaVision DaVis 7 was applied
to control all hardware components and for 32 bit image acquisition
and visualization. This software package comprises PIV software for
calculating the flow fields using cross-correlation analysis.

An incense smoke with sub-micron particles as a tracer was used for
the PIV measurements. Smoke was produced by high temperature
sublimation of solid incense particles. Analysis of smoke particles
using a microscope (Nikon, Epiphot with an amplification of 560) and
a PM-300 portable laser particulate analyzer showed that these
particles have an approximately spherical shape and that their mean
diameter is of the order of $0.7 \mu$m. The probability density
function of the particle size measured with the PM300 particulate
analyzer was independent of the location in the flow for incense
particle size of $0.5-1 \, \mu $m.

Series of 130 pairs of images acquired with a frequency of 4 Hz were
stored for calculating the velocity maps and for ensemble and
spatial averaging of turbulence characteristics. The center of the
probed flow region coincides with the center of the chamber. We
measured the velocity field in the flow areas $487.5 \times 240$
mm$^2$ with a spatial resolution of $234.7 \mu$m~/~pixel. These
regions were analyzed with interrogation windows of $32 \times 32$
pixels. A velocity vector was determined in every interrogation
window, allowing us to construct a velocity map comprising $65
\times 32$ vectors.

Mean and r.m.s. velocities, two-point correlation functions and an
integral scale of turbulence were determined from the measured
velocity fields. The mean and r.m.s. velocities for each point of
the velocity map (1024 points) were determined by averaging over 130
independent maps. The two-point correlation functions of the
velocity field were determined for each  point of the central part
of the velocity map ($32 \times 32$ vectors) by averaging over 130
independent velocity maps. An integral scale $l$ of turbulence was
determined from the two-point correlation functions of the velocity
field. These measurements were repeated for various temperature
differences between the bottom and the top walls of the chamber. The
size of the probed region did not affect our results. Similar
experimental set-up and data processing procedure were used in
experimental study by Eidelman et al. (2002; 2004) and Buchholz
et~al. (2004) of a new phenomenon of turbulent thermal diffusion
(see Elperin et al. 1996; 1997; 2001).

The maximum tracer particle displacement in the experiment was of
the order of $8$ pixels, i.e., $1/4$ of the interrogation window.
The average displacement of tracer particles was of the order of
$2.5$ pixels. Therefore, the average accuracy of the velocity
measurements was of the order of $4 \%$ for the accuracy of the
correlation peak detection in the interrogation window of the order
of $0.1$ pixels (see, e.g., Adrian 1991; Westerweel 1997; 2000).

\section{Hysteresis phenomenon}

\begin{figure}
\vspace*{2mm} \centering
\caption{\label{Fig3} Mean flow patterns during increase of the
the temperature difference between the bottom and the top  walls
of the chamber: (a) $\Delta T = 10$ K $({\rm Ra} = 0.18 \times
10^8)$; (b) $\Delta T = 20$ K $({\rm Ra} = 0.33 \times 10^8)$; (c)
$\Delta T = 70$ K $({\rm Ra} = 0.8 \times 10^8)$.}
\end{figure}

\begin{figure}
\vspace*{2mm} \centering
\caption{\label{Fig4} Mean flow patterns during decrease of the
the temperature difference between the bottom and the top walls of
the chamber: (a) $\Delta T = 60$ K $({\rm Ra} = 0.74 \times
10^8)$; (b) $\Delta T = 35$ K $({\rm Ra} = 0.52 \times 10^8)$; (c)
$\Delta T = 15$ K $({\rm Ra} = 0.26 \times 10^8)$.}
\end{figure}

In this Section we discuss the hysteresis phenomenon which was found
in the turbulent convection by varying the temperature difference
between the bottom and the top walls of the chamber within a range
of $5$ to $80$ K (the Rayleigh number was changed within the range
of ${\rm Ra} \approx 10^7 - 10^8$). The hysteresis loop comprises
the two-cell and one-cell flow patterns while the aspect ratio of
the chamber $A=2.23$ is kept constant. In particular, increasing the
temperature difference from $5$ to $15$~K we observed first two-cell
flow pattern with the downward motions in the central region of the
chamber between two cells (Fig. 3a), then one-cell flow pattern
within a range of the temperature difference of $20$ to $30$ K with
the counterclockwise mean flow (Fig. 3b). Further increase of the
temperature difference from $35$ to $80$ K results in two-cell flow
pattern with the upward mean flow in the central region of the
chamber between the two cells (Fig. 3c). Decreasing the temperature
difference we observed two-cell flow pattern within a range from
$80$ to $45$ K with the upward flow in the region between two cells
(Fig. 4a), then one-cell flow pattern with clockwise mean flow
within a range of the temperature difference of $40$ to $20$ K (Fig.
4b). Further decrease of the temperature difference from $15$ to $5$
K results in two-cell flow pattern with the downward mean flow in
the central region of the chamber between the two cells (Fig. 4c).

During the transition from the two-cell to one-cell flow patterns,
the direction of the mean flow in the one-cell pattern coincided
with that of the right cell in the two-cell flow pattern. The
latter may be caused by an asymmetry of the chamber due to the
presence of the grid near the left wall. This grid introduces an
additional anisotropy in the small-scale turbulent convection. In
all experiments the transition from the two-cell to one-cell flow
patterns was found to be accompanied by the change of the sign of
the degree of anisotropy $\chi$ of turbulent velocity field, which
is defined as
\begin{eqnarray}
\chi = {4 \over 3} \biggl[{\langle {\bf u}_{y}^{2} \rangle \over
\langle {\bf u}_{z}^{2} \rangle} - 1\biggr] \; .
\label{A6}
\end{eqnarray}
In particular, the dependence of the degree of anisotropy $\chi$ of
turbulent velocity field on the Rayleigh number obtained in the
experiment is shown in Fig.~5. The parameter $\chi$ was negative for
the two-cell flow pattern and was positive for the one-cell flow
pattern.

\begin{figure}
\vspace*{2mm} \centering
\includegraphics[width=8cm]{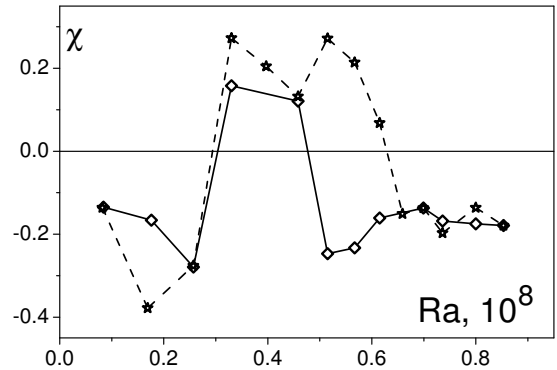}
\caption{\label{Fig5} Degree of anisotropy $\chi$ of turbulent
velocity field versus the Rayleigh number: increase of ${\rm Ra}$
(solid line) and decrease of ${\rm Ra}$ (dashed line).}
\end{figure}

These experimental observations can be explained by invoking the
theory of formation of coherent structures developed by Elperin et
al. (2002; 2005). In particular, Fig.~6 shows the theoretical
dependence of the growth rate of the large-scale instability
versus the parameter $\chi$ for the one-cell and two-cell flow
patterns. Note that the aspect ratio of the large-scale cell in
the two-cell flow pattern observed in the experiment was
approximately $1$ (see Figs.~3a; 3c; 4a; 4c), while in the
one-cell flow pattern it was approximately $2$ (see Figs.~3b; 4b).
The growth rate of the large-scale instability for the two-cell
mode, $\gamma_2$, is larger than that for the one-cell mode,
$\gamma_1$, for negative values of the degree of anisotropy $\chi$
of turbulent velocity field, and $\gamma_2 < \gamma_1$ for
positive values of the parameter $\chi$. Figure~6 is plotted for
the degree of thermal anisotropy $\alpha=1.7$.

\begin{figure}
\vspace*{2mm} \centering
\includegraphics[width=8cm]{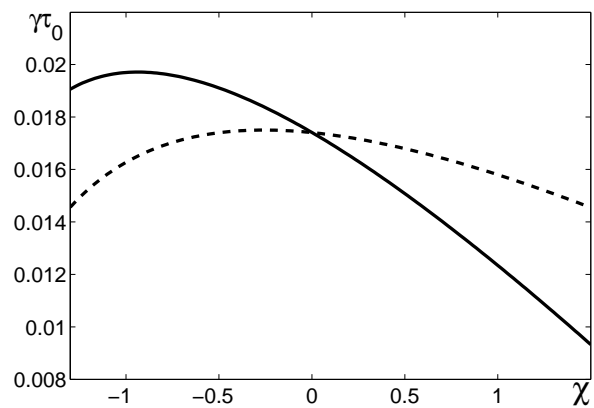}
\caption{\label{Fig6} Growth rate of the large-scale instability
versus the parameter $\chi$ for the one-cell flow pattern (dashed
line) and for the two-cell flow pattern (solid line), for
$\alpha=1.7$ and $L / l_{0}=10$.}
\end{figure}

\begin{figure}
\vspace*{2mm} \centering
\includegraphics[width=8cm]{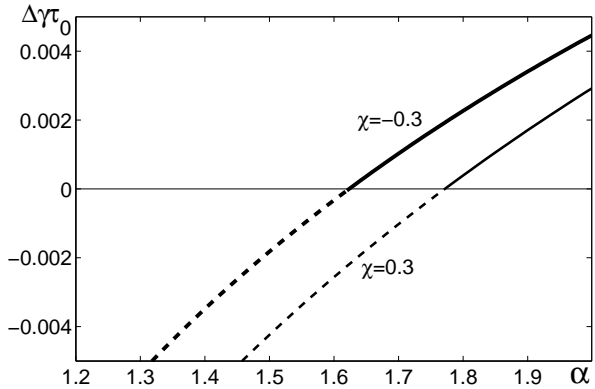}
\caption{\label{Fig7} The difference $\Delta \gamma = \gamma_2 -
\gamma_1$ in the growth rate of the large-scale instability for
the two-cell and one-cell flow patterns versus the degree of
thermal anisotropy $\alpha$ for different values of the parameter
$\chi$: $\quad \chi = - 0.3$ (thick curve) and $\chi = 0.3$ (thin
curve), for $L / l_{0}=10$. The range with $\Delta \gamma > 0$
corresponds to the two-cell flow pattern (dashed line), and the
range with $\Delta \gamma < 0$ is for the one-cell flow pattern
(solid line).}
\end{figure}

The difference $\Delta \gamma = \gamma_2 - \gamma_1$ of the growth
rates of the large-scale instability for the two-cell and one-cell
flow patterns versus the degree of thermal anisotropy $\alpha$ is
shown in Fig.~7. The range of parameters where $\Delta \gamma >
0$, corresponds to the two-cell flow pattern, since in this range
the growth rate of the excitation of the two-cell mode is larger
than that of the one-cell mode. Accordingly, the range of
parameters where $\Delta \gamma < 0$, corresponds to the one-cell
flow pattern, because in this range the growth rate of the
excitation of the two-cell mode is smaller than that of the
one-cell mode. Let us define the point $\chi=\chi_b$ for a given
parameter $\alpha$ where $\Delta \gamma = 0$.  The dependence of
the bifurcation point $\chi_b$ versus the degree of thermal
anisotropy $\alpha$ is shown in Fig.~8. The value $\chi_b$
increases with the increase of the parameter $\alpha$.

\begin{figure}
\vspace*{2mm} \centering
\includegraphics[width=8cm]{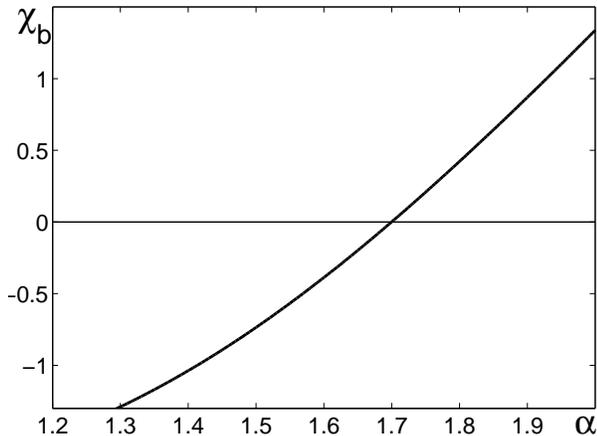}
\caption{\label{Fig8} Bifurcation point $\chi_b$ (where $\Delta
\gamma = 0)$ versus the degree of thermal anisotropy $\alpha$ for
$L / l_{0}=10$.}
\end{figure}

Note that the range of the degree of thermal anisotropy $\alpha$
when two kinds of flow patterns (the two-cell and one-cell flow
patterns) can exist in the system is very narrow. For instance, when
$-0.3 \leq \chi \leq 0.3$ the parameter $\alpha$ changes within the
range between $1.63$ and $1.77$ (see Fig.~7). The latter theoretical
prediction is in an agreement with the experimental observations. In
particular, our preliminary measurements show that the degree of
thermal anisotropy $\alpha$ in the chamber in the experiments was of
the order of $1 - 2$. Therefore, the above discussion shows that the
obtained experimental results are in a good agreement with the
predictions of the theory of formation of coherent structures in
turbulent convection (see Elperin et al. 2002; 2005).

A number of experiments were also conducted with different
additional thermal insulation of the side walls of the chamber in
order to study whether a heat flux through the side walls affects
the turbulent convective pattern. In particular, we plotted in
Fig.~9 the degree of anisotropy $\chi$ of turbulent velocity field
versus the Rayleigh number for the chamber with additional thermal
insolation with Perspex plates (dashed line) and for the experiments
without the additional thermal insolation (solid line) in the case
of increase of the Rayleigh number (Fig.~9a) and during decrease of
${\rm Ra}$ (Fig.~9b). The difference in the dependence $\chi({\rm
Ra})$ was not essential, and the hysteresis phenomenon was observed
in these experiments as well.

Notably, the vertical grid (see Fig.~2) which has been placed
asymmetrically, introduces additional anisotropy in turbulent fluid
flow in a chamber. Without this vertical grid the dynamics of flow
patterns becomes more complicated. In particular, after a two-cell
flow pattern at higher values of the Rayleigh number we often
observed an additional transition to the one-cell flow pattern.
However, we did observe the hysteresis phenomenon in this case. When
we also changed the aspect ratio from $A=2$ to $A=2.23$, the
hysteresis phenomenon was observed as well.

\begin{figure}
\vspace*{2mm} \centering
\includegraphics[width=8cm]{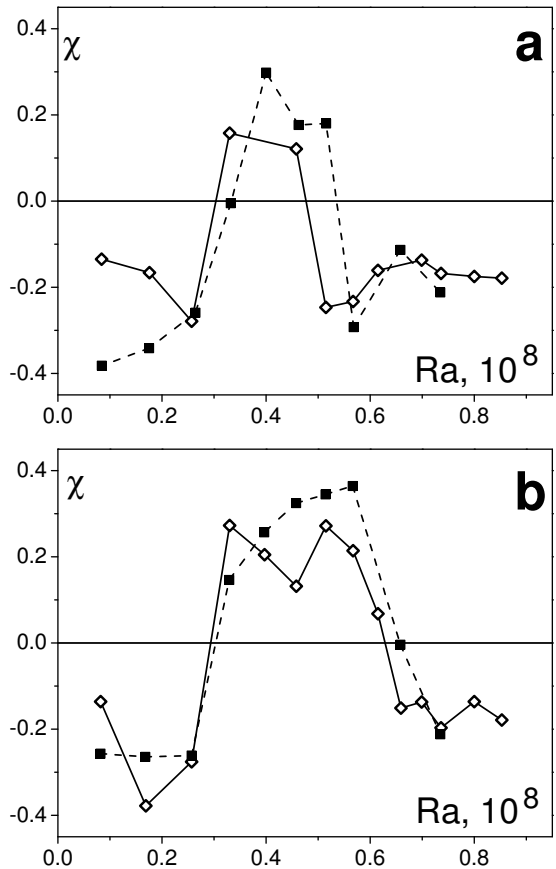}
\caption{\label{Fig9} Degree of anisotropy $\chi$ of turbulent
velocity field versus the Rayleigh number for the experiments shown
in Fig. 5 (solid line) and for the chamber with additional thermal
insolation (dashed line) during: (a) increase of ${\rm Ra}$ and (b)
decrease of ${\rm Ra}$.}
\end{figure}

\section{Conclusions}

In the present study we demonstrated that the anisotropy of
turbulent velocity field plays a crucial role in the hysteresis
phenomenon that was found in our experiments in turbulent
convection. The hysteresis loop comprises the two-cell and one-cell
flow patterns while the aspect ratio of the chamber is kept
constant. The observed transition from the two-cell to one-cell flow
patterns causes a drastic change of the degree of anisotropy $\chi$
of the turbulent velocity field from negative to positive values.
This finding is in a good agreement with the theoretical
predictions. In particular, we found that there is a critical value
$\chi_b$ of the degree of anisotropy of turbulent velocity field
when the growth rate of the large-scale instability for the one-cell
flow pattern is larger than that for the two-cell flow pattern. This
critical value depends on the degree of thermal anisotropy.
Transition through this critical value $\chi_b$ results in a
bifurcation, i.e., it causes the transition from the two-cell to
one-cell flow patterns. Then further increase of the temperature
difference between the bottom and the top walls of the chamber
decreases the degree of anisotropy $\chi$, and the system passes
again through the bifurcation from one-cell to two-cell flow
patterns. Decreasing the temperature difference causes these
bifurcations but for different values of the temperature difference.
Therefore, we observed the hysteresis phenomenon in turbulent
convection. The developed theory of coherent large-scale
circulations in turbulent convection (Elperin et al. 2002; 2005) is
in a good agreement with the experimental observations.

Note that the nonlinear dynamics may play a very essential role in
the observed transitions between one-cell flow pattern to the
two-cell flow pattern (and vice versa) when we change the values of
the Rayleigh numbers. After parametrization of the turbulence the
nonlinear mean-field equations determine the dynamics of modes (flow
patterns). In our experiments with $A=2 - 2.23$ we have a two-mode
system (one-cell and two-cell modes), and nonlinear interactions
between the modes play a very essential role. Experiments and
numerical simulations with large aspect ratios (Willis et al. 1972;
Fitzjarrald 1976; Hartlep et al. 2003) can be regarded as a
multi-mode system, which may be strongly different from the two-mode
system.

The obtained results might be important in atmospheric turbulent
convection (see, e.g., Zilitinkevich 1991; Etling and Brown 1993;
Atkinson and Wu Zhang 1996; Br\"{u}mmer 1999) and industrial
turbulent flows (see, e.g., Kakac et al. 1987; Incropera and DeWitt
2002).

\bigskip
\begin{acknowledgments}
The authors benefited from stimulating discussions with F.~Busse,
S.~Grossmann, S.~Morris, A.~Tsinober, X.Q.~Xia and V.~Yakhot. We
thank A.~Krein for his assistance in construction of the
experimental set-up and performing the experiments. We also thank
I.~Golubev for his assistance in processing the experimental data.
This work was partially supported by the German-Israeli Project
Cooperation (DIP) administrated by the Federal Ministry for
Education and Research (BMBF) and by the Israel Science Foundation
governed by the Israeli Academy of Science.
\end{acknowledgments}


\begin{thebibliography}{}

\bibitem {A91}
Adrian RJ (1991) Particle imaging techniques for experimental fluid
mechanics. Ann Rev Fluid Mech 23: 261-304

\bibitem {AZ96}
Atkinson BW; Wu Zhang J (1996) Mesoscale shallow convection in the
atmosphere. Rev Geophys 34: 403-431

\bibitem {BM96}
Braunsfurth MG; Mullin T (1996) An experimental study of oscillatory
convection in liquid gallium. J Fluid Mech 327: 199-219

\bibitem {BN05}
Brown E; Nikolaenko A; Ahlers G (2005) Orientation changes of the
large-scale circulation in turbulent Rayleigh-B{\'e}nard convection,
Phys Rev Lett 95: 084503 (1-4)

\bibitem {B99}
Br\"{u}mmer B (1999) Roll and cell convection in winter-time
arctic cold-air outbreaks. J Atmosph Sci 56: 2613-2636

\bibitem {BE04}
Buchholz J; Eidelman A; Elperin T; Gr\"{u}nefeld G; Kleeorin N;
Krein A; Rogachevskii I (2004) Experimental study of turbulent
thermal diffusion in oscillating grids turbulence. Exp Fluids 36:
879-887

\bibitem {BK03}
Burr U; Kinzelbach W; Tsinober A (2003) Is the turbulent wind in
convective flows driven by fluctuations? Phys Fluids 15: 2313-2320

\bibitem {B67}
Busse FH (1967) The stability of finite amplitude cellular
convection and its relation to an extremum principle. J Fluid Mech
30: 625-649

\bibitem {B78}
Busse FH (1978) Non-linear properties of thermal convection. Rep
Prog Phys 41: 1929-1967

\bibitem {BW71}
Busse FH and Whitehead JA (1971) Instabilities of convective rolls
in a high Prandtl number fluid. J Fluid Mech 47: 305-320

\bibitem {BW74}
Busse FH and Whitehead JA (1974) Oscillatory and collective
instabilities rolls in large Reynolds number convection. J Fluid
Mech 66: 67-79

\bibitem {B83}
Busse FH (1983) Generation of mean flows by thermal convection.
Physica D 9: 287-299

\bibitem {CC96}
Ciliberto S; Cioni S; Laroche C (1996) Large-scale flow properties
of turbulent thermal convection. Phys Rev E 54: R5901-R5904

\bibitem {EEKR02}
Eidelman A; Elperin T; Kapusta A; Kleeorin N; Krein A; Rogachevskii
I (2002) Oscillating grids turbulence generator for turbulent
transport studies. Nonlinear Proc Geoph 9: 201-205

\bibitem {EEKR04}
Eidelman A; Elperin T; Kleeorin N; Krein A; Rogachevskii I; Buchholz
J; Gr\"{u}nefeld G (2004) Turbulent thermal diffusion of aerosols in
geophysics and in laboratory experiments. Nonlinear Proc Geoph 11:
343-350

\bibitem {EKR96}
Elperin T; Kleeorin N; Rogachevskii I (1996) Turbulent thermal
diffision of small inertial particles. Phys Rev Lett 76: 224-228

\bibitem {EKR97}
Elperin T; Kleeorin N; Rogachevskii I (1997) Turbulent
barodiffusion, turbulent thermal diffusion and large-scale
instability in gases. Phys Rev E 55: 2713-2721

\bibitem {EKR01}
Elperin T; Kleeorin N; Rogachevskii I; Sokoloff D (2001) Mean-field
theory for a passive scalar advected by a turbulent velocity field
with a random renewal time. Phys Rev E 64: 026304 (1-9)

\bibitem {EKR02}
Elperin T; Kleeorin N; Rogachevskii I; Zilitinkevich S  (2002)
Formation of large-scale semiorganized structures in turbulent
convection. Phys Rev E 66: 066305 (1-15)

\bibitem {EKR05}
Elperin T; Kleeorin N; Rogachevskii I; Zilitinkevich S (2005)
Tangling turbulence and semi-organized structures in convective
boundary layers. Boundary-Layer Meteorol, accepted

\bibitem {EB93}
Etling D; Brown RA (1993) Roll vortices in the planetary boundary
layer: a review. Boundary-Layer Meteorol 65: 215-248

\bibitem {F76}
Fitzjarrald DE (1976) An experimental study of turbulent convection
in air. J Fluid Mech 73: 693-719

\bibitem {GB99}
Gelfgat AYu; Bar-Yoseph PZ; Yarin AL (1999) Stability of multiple
steady states of convection in laterally heated cavities. J Fluid
Mech 388: 315-334

\bibitem {HT03}
Hartlep T; Tilgner A; and Busse FH (2003) Large-scale structures in
Rayleigh-Benard convection at high Rayleigh numbers. Phys Rev Lett
91: 064501 (1-4)

\bibitem {H84}
Hunt JCR (1984) Turbulence structure in thermal convection and
shear-free boundary layers. J Fluid Mech 138: 161-184

\bibitem {HK88}
Hunt JCR; Kaimal JC; Gaynor JI (1988) Eddy structure in the
convective boundary layer - new measurements and new concepts. Quart
J Roy Meteorol Soc 114: 837-858

\bibitem {ID02}
Incropera FP; DeWitt DP (2002) Fundamentals of Heat and Mass
Transfer. New York

\bibitem {IY02}
Ishihara T; Yoshida K; Kaneda Y (2002) Anisotropic velocity
correlation spectrum at small scales in a  homogeneous turbulent
shear flow. Phys. Rev. Lett. 88: 154501 (1-4)

\bibitem {K01}
Kadanoff LP (2001) Turbulent heat flow: structures and scaling.
Phys. Today 54: 34-38

\bibitem {KR87}
Kakac S; Ramesh SK; Aung W (1987) Handbook of Single-Phase
Convective Heat Transfer. New York

\bibitem {KH81}
Krishnamurti R; Howard LN (1981) Large-scale flow generation in
turbulent convection. Proc Natl Acad Sci USA 78: 1981-1985

\bibitem {LS80}
Lenschow DH; Stephens PL (1980) The role of thermals in the
convective boundary layer. Boundary-Layer Meteorol. 19: 509-532

\bibitem {L67}
Lumley JL (1967) Rational approach to relations between motions of
different scales in turbulent flows. Phys Fluids 10: 1405-1408

\bibitem {MY75}
Monin AS; Yaglom AM (1975) Statistical Fluid Mechanics. Cambridge,
Massachusetts, MIT Press. Vol. 2

\bibitem {NS01}
Niemela JJ; Skrbek L; Sreenivasan KR; Donnelly RJ (2001) The wind in
confined thermal convection. J Fluid Mech 449: 169-178

\bibitem {NS03}
Niemela JJ; Sreenivasan KR (2003) Rayleigh-number evolution of
large-scale coherent motion in turbulent convection. Europhys Lett
62: 829-833

\bibitem {PH04}
Parodi A; von Hardenberg J; Passoni G; Provenzale A; Spiegel EA
(2004) Clustering of plumes in turbulent convection. Phys Rev Lett
92: 194503 (1-4)

\bibitem {RW98}
Raffel M; Willert C; Kompenhans J (1998) Particle Image Velocimetry.
Springer

\bibitem {SV94}
Saddoughi SG; Veeravalli SV (1994) Local isotropy in turbulent
boundary layers at high Reynolds number. J Fluid Mech 268: 333-347

\bibitem {SS89}
Schmidt H; Schumann U (1989) Coherent structure in the convective
boundary layer derived from large-eddy simulations. J Fluid Mech
200: 511-562

\bibitem {S94}
Siggia ED (1994) High Rayleigh number convection. Annu Rev Fluid
Mech 26: 137-168

\bibitem {SQ04}
Shang XD; Qiu XL; Tong P; Xia XQ (2004) Measurements of the local
convective heat flux in turbulent Rayleigh-Benard convection. Phys
Rev E 70: 026308 (1-13)

\bibitem {SB02}
Sreenivasan KR; Bershadskii A; Niemela JJ (2002) Mean wind and its
reversal in thermal convection. Phys Rev E 65: 056306 (1-11)

\bibitem {TM05}
Tsuji Y; Mizuno T; Mashiko T; Sano M (2005) Mean wind in convective
turbulence of Mercury. Phys Rev Lett 94: 034501 (1-4)

\bibitem {W97}
Westerweel J (1997) Fundamentals of digital particle image
velocimetry. Measurem Sci Technology 8: 1379-1392


\bibitem {W00}
Westerweel J (2000) Theoretical analysis of the measurement
precision of particle image velocimetry. Exp Fluids Suppl 29: S3-S12

\bibitem {WH92}
Williams AG; Hacker JM (1992) The composite shape and structure of
coherent eddies in the convective boundary layer. Boundary-Layer
Meteorol 61: 213-245

\bibitem {WH93}
Williams AG; Hacker JM (1993) Interaction between coherent eddies in
the lower convective boundary layer. Boundary-Layer Meteorol 64:
55-74

\bibitem {WD72}
Willis GE; Deardorff JW; Somerville RCJ (1972) Roll-diameter
dependence in Rayleigh convection and its effect upon the heat flux.
J Fluid Mech 54: 351-367

\bibitem {WC72}
Wyngaard JC; Cote OR (1972) Cospectral similarity in the atmospheric
surface layer. Quart J Royal Meteorol Soc 98: 590-602

\bibitem {XL04}
Xi HD; Lam S; Xia XQ (2004) From laminar plumes to organized flows:
the onset of large-scale circulation in turbulent thermal
convection. J Fluid Mech 503: 47-56

\bibitem {YK02}
Young GS; Kristovich DAR; Hjelmfelt MR; Foster RC (2002) Rolls,
streets, waves and more. Bull Amer Meteorol Soc 83: 997-1001

\bibitem {YI03}
Yoshida K; Ishihara T; Kaneda Y (2003) Anisotropic spectrum of
homogeneous turbulent shear flow in a Lagrangian renormalized
approximation. Phys Fluids 15: 2385-2397

\bibitem {Z91}
Zilitinkevich SS (1991) {\em Turbulent Penetrative Convection},
Avebury Technical, Aldershot

\bibitem {ZG98}
Zilitinkevich SS; Grachev A; Hunt JCR (1998) Surface frictional
processes and non-local heat/mass transfer in the shear-free
convective boundary layer. In Plate et al. (eds.), Buoyant
convection in geophysical flows. Kluwer Academic Publications,
Dordrecht, The Netherlands, 83-113

\bibitem {ZM90}
Zocchi G; Moses E; Libchaber A (1990) Coherent structures in
turbulent convection. Physica A 166: 387-407

\end{thebibliography}
\end{document}